# Percolation theory suggests some general features in range margins across environmental gradients

Róbert Juhász[1] and Beáta Oborny[2,3]*

[1] Department of Theoretical Solid State Physics, Wigner Research Centre for Physics. Address: 29-33 Konkoly-Thege Miklós út, Budapest, Hungary, H-1121.
E-mail: juhasz.robert@wigner.mta.hu
ORCID ID: 0000-0003-2617-5266

[2] Department of Plant Taxonomy, Ecology and Theoretical Biology, Eötvös Loránd University. Address: 1/C Pázmány Péter stny, Budapest, Hungary, H-1117.
E-mail: beata.oborny@ttk.elte.hu
ORCID ID: 0000-0003-2997-9921

[3] GINOP Sustainable Ecosystems Group, MTA Centre for Ecological Research.
Address: 3 Klebelsberg Kuno u., Tihany, H-8237.

*Corresponding author.

Abstract

The margins within the geographic range of species are often specific in terms of ecological and evolutionary processes, and can strongly influence the species' reaction to climate change. One of the frequently observed features at range margins is fragmentation, caused internally by population dynamics or externally by the limited availability of suitable habitat sites. We study both causes, and describe the transition from a connected to a fragmented state across space using a gradient metapopulation model. Our approach is characterized by the following features. 1) Inhomogeneities can occur at two spatial scales: there is a broad-scale gradient, which can be patterned by fine-scale heterogeneities. 2) We study the occupancy of this terrain in a steady-state on two temporal scales: in snapshots and by long-term averages. The simulations reveal some general scaling laws that are applicable in various environments, independently of the mechanism of fragmentation. The edge of the connected region (the hull) is a fractal with dimension 7/4. Its width and length changes with the gradient according to universal scaling laws, that are characteristic for the percolation transition. The results suggest that percolation theory is a powerful tool for understanding the structure of range margins in a broad variety of real-life scenarios, including those in which the environmental gradient is combined with fine-scale heterogeneity. This provides a new method for comparing the range margins of different species in various geographic regions, and monitoring range shifts under climate change.

Keywords:

metapopulation dynamics, habitat loss and fragmentation, species borders, core and peripheral populations, distributional limits, contact process

Introduction

The margins of geographic ranges have been in the focus of ecology and evolutionary biology for several decades. In comparison with the core of distribution, they are often specific in terms of genetic composition and in the manifestation of phenotypic plasticity (Holt and Keitt 2005; Gaston 2009; Kubisch et al. 2014). Typically, an environmental gradient causes a decline in the abundance of a species (Brown et al. 1996; Vucetich and Waite 2003). Thus, the response of the population to the key environmental factor(s) can directly be observed across space (altitude, latitude or other direction). The study of range margins helps to understand the natural limitations of a species, and to predict the direction and magnitude of range shift in case of a climate change (Travis 2003; Best et al. 2007; Geber 2008; Mustin et al. 2009; Turner and Wong 2010; Eppinga et al. 2013; Tejo et al. 2017). Usually multiple environmental factors are involved (temperature, moisture, etc.).

Metapopulation modelling simplifies these complex scenarios by considering the changes in the demographic rates across space (Wilson et al. 1996; Holt and Keitt 2000; Gastner et al. 2009; Oborny et al. 2009). Thus, it provides a unified theoretical framework for the study of range dynamics across environmental gradients [see Holt et al. (2005); Kubisch et al. (2014) and Oborny (2018) for reviews].

The most fundamental variables are the rate of colonization ($c$), and the rate of local extinction ($e$) in each habitat site. In the simplest gradient situation, c and/or $1/e$ changes linearly along a spatial axis $x$. (The choice of $1/e$ instead of $e$ is motivated by using $\lambda = \frac{c}{e}$ as a control parameter; see later). We refer to this case as the *even slope* (ES) model (**Fig. 1.a**). In more complex cases, further heterogeneity may occur beside the gradient. For example, a hillside is usually uneven topographically, and it may contain rock outcrops. To study these cases, we introduce a *rugged slope* (RS) model, in which fine-scale heterogeneity also occurs (**Fig. 1.b**).

Gradient metapopulation models that can be classified as ES models have been used from the 1990s (Lennon et al. 1997; Holt and Keitt 2000; Holt et al. 2005; Antonovics et al. 2006; Gastner et al. 2009; Oborny et al. 2009). Although the details differed in these models, they share two main conclusions. 1) A smooth change in the environment causes an abrupt change in the steady-state occupancy *n(x)*. 2) Towards the range limit, the originally continuous occupancy gets fragmented. These conclusions are common with many other kinds of gradient metapopulation models as well [e.g., Wilson et al. (1996); Zeng and Malanson (2006); see Oborny (2018) for a review].

To complement these approaches, some range edge models have included fine-scale heterogeneity, but without any gradient in the demographic rates (Holt and Keitt 2000; Holt et al. 2005). Our RS model and a patch occupancy model introduced by Mustin et al. (2009) combine these approaches, as they include a demographic gradient as well as fine-scale heterogeneity. In the model of Mustin et al., the primary focus was on the density profile of the whole metapopulation, and on the distortion of this profile during a range shift. In the present study, there is no range shift, and we investigate the two-dimensional pattern of occupancy.

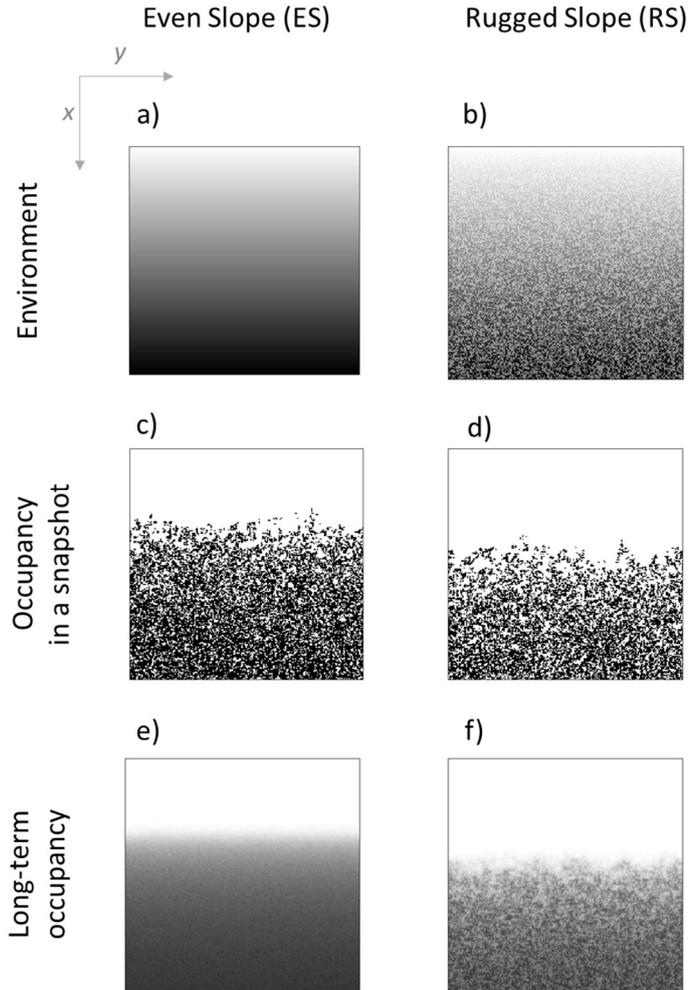

***Fig. 1.*** *The environment and the steady-state occupancy in a lattice with L=200. **a)** The environmental gradient in the ES model. **b)** The same gradient with fine-scale heterogeneity in the RS model. Shading indicates the quality of the environment from c=0 (white) to c=1 (black). The abundance of bad sites is a=0.5, and their penetrability is w=0.5. (See the definition of a and w in the Methods.) **c)** and **d)** Snapshots showing the occupancies at time $2^{14}$ (measured in Monte Carlo steps). White/black denote empty/occupied. **e)** and **f)** Long-term averages of the occupancy, measured after time $2^{14}$ (see the text for details). Shading shows the occupancies from 0 (white) to 1 (black).*

It has been generally observed in the above-mentioned gradient metapopulation models that the range margin consisted of a 'mainland' (at the favourable end of the slope) and of several 'islands' (at the unfavourable one). Real-life range margins have also been observed to get fragmented towards the edge (Brown et al. 1996; Zeng and Malanson 2006; Kunin et al. 2009; Eppinga et al. 2013; Saravia et al. 2018). Milne et al. (1996) separated the connected from the fragmented portion of the occupied sites, and delineated the hull of the connected patch in a pinyon-juniper woodland in New Mexico. To our knowledge, they were the first researchers who proposed the use of percolation theory for analyzing range margins. We continued these investigations (Gastner et al. 2009) utilizing that percolation theory suggests some universal scaling laws concerning the hull (see these laws in the Methods). Their validity was checked in simulated patterns produced by an ES model; and the applicability of the method was demonstrated on a satellite image of another pinyon-juniper woodland in New Mexico (Gastner et al. 2009).

The present paper focuses on the effect of fine-grained heterogeneity in the environment. In the RS model, we place small obstacles onto the slope, and ask the following questions.
1) How does the abundance and penetrability of the obstacles influence the range edge?
2) Are the scaling laws detected in snapshots also applicable in the long run, for the averages of occupancies?

We show that the scaling laws are broadly applicable. Accordingly, we propose that the geographic distributions of species should be delineated at the hull in order to monitor range shifts reliably.

## Methods

*A metapopulation model across an environmental gradient*

The model is based on the so-called contact process (CP), which is a frequently applied tool for modelling the occupancy of space by metapopulations [see Oborny et al. (2007) for review]. It is typically implemented in regular lattices, in which every lattice cell (site) has $z$ neighbours. A site can be empty or occupied. The following transitions occur independently. An empty site $i$ can become occupied by colonization from the neighborhood. If $h_i$ of the $z$ neighbors are occupied, then the rate of occupation is $c_i \cdot \frac{h_i}{z}$, where $c_i$ is the local colonization rate. Note that the rate of colonization from a donor site to a recipient one is determined in this formulation by a parameter ($c_i$) characteristic of the recipient site. Occupied cells can become empty spontaneously with a local extinction rate $e_i$.

In the simplest, homogeneous CP (Harris 1974), the values of $c$ and $e$ are constant, i.e., uniform in every site. Time can thus be conveniently rescaled by the average lifetime of occupancy in each cell, $1/e$. Therefore, the proportion of occupied sites ($n$) in the steady-state depends only on a single parameter, $\lambda = \frac{c}{e}$. Changing this control parameter, $\lambda$, the metapopulation undergoes a phase transition from a living to an extinct state at a critical threshold $\lambda_c$, where the correlation length is infinite (Marro et al. 2005).

In the even slope (ES) contact processes, the demographic parameters change smoothly along a spatial axis ($x$); while they are constant in the perpendicular, $y$ direction (**Fig. 1.a**). According to earlier results with these *c(x)* and *e(x)* functions, $\lambda(x) = \frac{c(x)}{e(x)}$ is a good predictor of the local density *n(x)*, i.e., it determines *n(x)* with high accuracy (Oborny et al. 2009).

In our ES and RS models, we have a two-dimensional square lattice of size $L \times L$, with open boundaries in the $x$ direction and periodic boundary conditions in the $y$ direction. The neighborhood of each lattice cell (site) consists of the four adjacent cells ($z$=4), and represents the dispersal kernel of the species. Thus, the spatial unit is defined according to the dispersal distance. $e$ is kept constant throughout the lattice. The environmental gradient is implemented by changing the colonizaton rate linearly in space as $c(x) = gx$, where $g$ denotes the gradient. We set $g = 1/L$. Therefore, c(x) always ranges from 1/*L* to 1.

In the RS model, we dilute the lattice by obstacles against spreading. The probability of a site being bad (an obstacle) is *a*, while being good (no obstacle) is 1-*a*. Good and bad sites are uniformly and independently distributed over the lattice. Good sites are colonized as described in the ES model: the local colonization rate is determined by the gradient, $\zeta(x,y) = c(x)$. Bad sites are colonized with a rate $\zeta(x,y) = w \cdot c(x)$, where $0 \leq w \leq 1$. Therefore, the ES model is a special case of the RS model, at *w*=1. In the forthcoming text, we refer to *a* as the *abundance* and to *w* as the *penetrability* of the obstacles. We assume that the spatial configuration of the obstacles does not change over the time span of the study. The best examples for such permanent, fine scale heterogeneities are microtopographic differences over a terrain, which can cause differences in the availability of water, soil quality, and other environmental variables. In statistical physics, this is referred to as quenched disorder. **Fig. 1.b** shows an example for the RS environment. For the dynamics of the contact process in the presence of quenched disorder, but without any gradient, we refer the reader to Moreira and Dickman (1996).

*Numerical simulations*

We performed numerical simulations and, after the system relaxed to the steady-state, we analyzed the occupancy of sites. **Figs. 1.c** and **d** present snapshots from the simulations in the ES and RS model, respectively. **Figs. 1.e** and **f** show long-term averages, produced by the state of each site. Using the same simulated environment, we let the metapopulation grow and reach a steady-state at time $T_1$. We then take snapshots of the configuration of the system periodically, separated by a time interval $T_2$. Let $q_{i,n}$ denote the state of site $i$ at the $n$th measurement. $q_{i,n}=0$ if the site is empty, and 1 if the site is occupied. The long-term average occupancy in site $i$ is defined as

$$\bar{n}_i = \frac{\sum_{n=1}^{N} q_{i,n}}{N},$$  Eqn. 1

where N denotes the number of measurements. In the simulations producing **Figs. 1.e** and **f**, the following parameter values were used: $T_1=2^{14}$ Monte Carlo steps (MCS), $T_2=2^{11}$ MCS, and N=1000. To study the scaling laws that characterize the hull (see later), we used $T_1=2^{15}$ MCS, $T_2=2^{12}$ MCS, and N=100. We ensured that the relaxation time $T_1$ was sufficient, checking that an increase of $T_1$ did not change the measured quantities significantly.

Note that, in the limit $N \to \infty$, the long-term average occupancies can be interpreted as local (site-dependent) occupation probabilities.

The study of the scaling laws necessitated to transform $\bar{n}_i$ into a binary variable. We considered the species present in site $i$, if its long-term occupancy exceeded an arbitrary threshold, $\bar{n}_i > b$, and it was considered absent otherwise. The occupancy threshold $b$ was 0.25, 0.50, or 0.75. Altogether, we analysed three kinds of patterns:

1) snapshot in the ES (**Fig. 1.c**),
2) snapshot in the RS (**Fig. 1.d**),
3) and the long-term occupancy in the RS (**Fig. 1.f**) in a binary version.

*Percolation structure of the range margin*

The above-mentioned spatial patterns can be analyzed in the framework of percolation theory. It is a powerful tool for revealing the connectivity structure within a set of sites (Stauffer and Aharony 1994); for ecological applications, see Loehle (1996), Milne et al. (1996), Li (2001), Oborny et al. (2007), Solé (2011) and the references cited therein. In the present model, let us define two occupied sites connected if, and only if, they are neighbours in the lattice (thus, we use the same neighbourhood size as in the contact process, although this is not a necessary choice). A set of directly or indirectly linked occupied sites makes a percolation cluster. Therefore, each percolation cluster is surrounded by unoccupied sites.

Traditional percolation models assume homogeneous space (no gradient), in which the occupied sites (called 'open' sites in the original terminology) are randomly distributed, and do not change over time. Let $p$ denote the probability that a site is occupied. Let us consider first a finite system: a lattice with size $L \times L$. It is an exciting question whether there exists a pathway along which it is possible to walk from one side of the lattice to the opposite side by stepping into occupied sites only. A cluster which contains such a pathway is referred to as a spanning cluster, and the probability of existence of a spanning cluster is called as percolation probability. It has been proven that the number of spanning clusters can only be 0 or 1 in the infinite system limit (Stauffer and Aharony 1994). Percolation theory has revealed that in an infinite system, there is a threshold in $p$ at which a spanning cluster emerges; i.e., the percolation probability $R=0$ at $p < p_c$, and $R>0$ at $p > p_c$, where $p_c$ is the percolation threshold. Numerical simulations have estimated the percolation threshold $p_c = 0.592746$ in the two-dimensional square lattice with four-cell neighborhood (Stauffer and Aharony 1994).

The shape of $R(p)$ is well-known in the vicinity of the threshold, at $p > p_c$.

$$R \propto (p - p_c)^\beta,  \quad \text{Eqn 2.}$$

where $\propto$ denotes 'proportional to'. The scaling exponent $\beta = 5/36$ is universal in the sense that it does not depend on the local details of the system, for example, on the lattice geometry or neighborhood size. It is fully determined by the dimensionality of the system (D=2 in the present case). This and other scaling laws in the vicinity of the threshold [see more in (Stauffer and Aharony 1994)] show that the system undergoes a continuous phase transition at the threshold. The universal nature of the scaling relations is related to the fact that the characteristic size of percolation clusters becomes infinite at the percolation threshold, which make the local details irrelevant. Among the universality classes of critical phenomena, this system belongs to the so-called isotropic percolation universality class [Bunde (1991); from a biological perspective, see Oborny et al. (2007)].

The pattern of occupied and empty sites in the steady state of our model differs from a standard percolation pattern in several respects. Let us discuss these differences, and consider first the contact process without a gradient.

First, the model assumes limited dispersal distance, as it states that colonization from an occupied site can only occur to neighboring empty sites. This induces spatial correlations between the occupancies of different sites. They are characterized by a correlation length $\xi$, which is finite apart from the critical threshold. Irrespective of the presence of these correlations, it is still possible to delineate percolation clusters and observe a phase transition, manifested by the appearance of a spanning cluster. Due to the spatial correlations, the threshold value of this transition is different from that of standard percolation (Gastner et al. 2009). We will return to the role of correlations in detail in the Discussion.

Second, our model contains an environmental gradient, which makes the occupancy change along $x$. According to the definition of the model, the average long-term occupancy of sites having the same $x$ coordinate is an increasing function of $x$. In the simulations, we have chosen the extinction rate in such a way that the occupancy near the edge $x=1$ is well below, while near the edge $x=L$ is well above the percolation threshold. As a consequence, at some coordinate in the interior of the lattice, the occupancy crosses its percolation threshold value.

We consider the largest percolation cluster present in the system, which we also refer to as the 'mainland', while other clusters are referred to as 'islands'. Similarly, we can define the largest cluster of unoccupied sites as the 'sea'. Due to our choice of the parameters, there are some properties in almost all samples in the large system size limit, i.e., the fraction of such samples tends rapidly to 1 as $L \to \infty$. These properties are the following. i) The mainland is present in the $x=L$ edge of the lattice, while absent in the $x=1$ edge; furthermore, the mainland is spanning in the $y$ direction, i.e., it has constituents at all transversal coordinates $y = 1, \dots, L$. ii) Similarly, the sea is present at $x = 1$, absent at $x = L$, and is spanning in the $y$ direction. iii) The mainland and the sea have a common interface, or, in other words, no more than one spanning cluster of occupied sites exists.

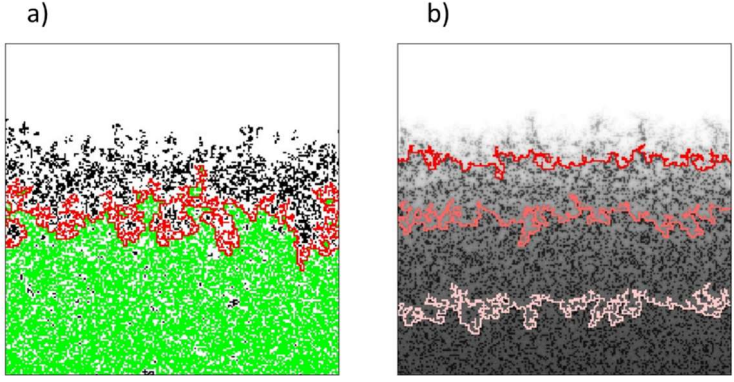

**Fig. 2.** *Connectivity of the occupied sites. Black: islands, red and green: mainland. Red marks the hull. The white sites are unoccupied. Note that white 'lakes' can occur even within the mainland, and these can also contain islands. The examples are the same as in Fig. 1.* ***a)*** *A snapshot in the ES model.* ***b)*** *Long-term average occupancies in the same realization. The occupancy thresholds are b=0.25, 0.50, and 2/3 respectively, from the top to the bottom of the figure.*

*The hull*

Within this complex structure, the hull of the mainland is one of the most remarkable objects (**Fig. 2.a)**. It can be defined as the 'coastline', i.e., the set of sites which belong to the mainland and are adjacent to the sea. In the limit of infinite system size, the hull is unique. In the simulations, we always checked whether the mainland and the sea spanned across the lattice in the *y* direction. We used only those realizations in which this condition was satisfied.

It has been numerically demonstrated in the ES (Gastner et al. 2009) that the hull is a fractal, and has the same fractal dimension, $d_f = 7/4$, as the hull of the spanning cluster in a similar non-gradient system [i.e., in ordinary percolation (Voss 1984; Saleur and Duplantier 1987)]. In this paper, we examine this in the RS as well, both in snapshots and in the binary patterns obtained from the long-term average occupancies (see above). We refer to the first as the *snapshot hull*, and to the second as the *long-term hull*.

In the simulated patterns, the mainland is determined by the Hoshen-Kopelman algorithm, and its hull is delineated by the left-turning biased walk method (Gastner et al. 2009). This method searches for the outermost path on the mainland, which is a sequence of sites visited by a left-turning biased walk. Due to the periodic boundary condition in the *y* direction, the path is closed, i.e. the walk returns to the starting position.

The length of the hull ($u$) is defined as the number of steps in the walk. The width $v$ of the hull is defined as the standard deviation of the *x* coordinates of the sites constituting the hull [as in Gastner et al. (2009)]: $v^2 = \frac{1}{u}\sum_{i=1}^{u}(x_i - \bar{x})^2$, where $\bar{x} = \frac{1}{u}\sum_{i=1}^{u} x_i$. In the numerical simulations, we calculated the average $u$ and $v$ over many independent random samples. In the case of the snapshot hull, the number of samples was 1000 for each size, except for the largest sizes, L=1024 and 2048, where it was 100 and 10, respectively. For the long-term hull, it was 100 for each size, except for L=2048, where it was 10. We investigate the dependence of the length ($u$) and width ($v$) of the hull on the environmental gradient *g*. The fractal dimension of the hull is estimated by the equipaced polygon method (Batty and Longley 1994; Gastner et al. 2009). For each hull, i.e. sequence of sites, we calculate the mean Euclidean distance $\overline{d(k)}$ between all pairs of sites separated by *k* steps in the sequence. The fractal dimension, $d_f$, can be then determined from the following scaling relation,

$$\overline{d(k)} \sim k^{1/d_f}. \qquad \text{Eqn. 3}$$

Results

The simulations show that the hull is a fractal in a broad range of $k$. This is the range within which a straight line can be fitted to the datapoints in **Fig. 3** (according to Eqn. 3). The fractal dimension is $d_f=7/4$ in every case.

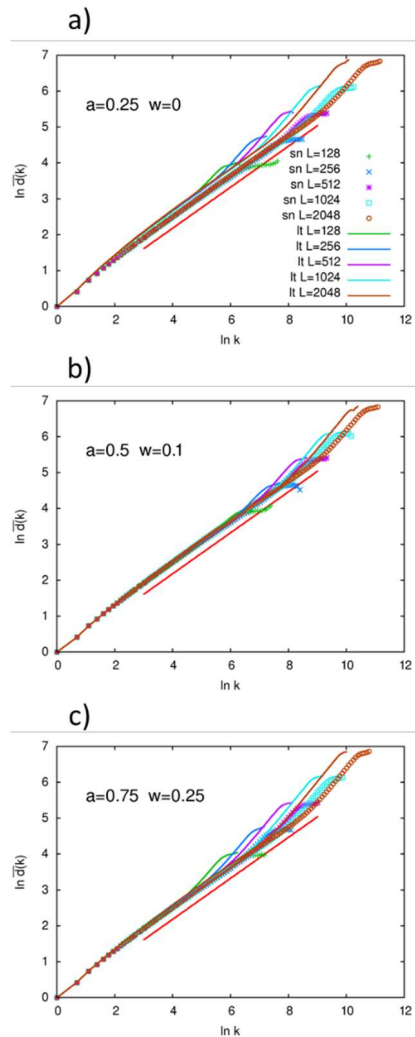

**Fig. 3.** *Estimation of the fractal dimension of the hull (according to Eqn 3) in the RS model. The simulations differ in the duration of the observation (snapshot versus long-term, abbreviated as sn versus lt), and in the system size (L). The legends in panel **a)** apply for all panels. The abundance (a) and penetrability (w) of the bad sites are indicated in the upper left corner. The occupancy threshold in the lt simulations is b=0.5. The straight line has a slope 4/7, which corresponds to a fractal dimension 7/4.*

The most notable result is that fine-grained heterogeneity did not influence the fractal dimension: it was 7/4 both in the ES (Gastner et al. 2009) and in the RS (**Fig. 3**), in spite of the fact that it did influence the mean position and width of the hull (compare **Figs. 1.c** and **d**). Varying the abundance (*a*) and penetrability (*w*) of the obstacles within the RS did not influence the fractal dimension either (compare **Figs. 3.a-c**). We varied *a* and *w* arbitrarily in rather broad ranges. In each case, the extinction rate was set so that the mean coordinate *x* of the hull is roughly in the middle of the lattice. With this choice, the survival of the population up to the end of the simulation was practically sure. **Fig. 3.a** was obtained in an environment in which the obstacles were sparse (*a*=0.25), and unpenetrable (*w*=0). Conversely, the habitat in **Fig. 3.c** contained a high number of penetrable obstacles (*a*=0.75, *w*=0.25). **Fig. 3.b** is intermediate. In spite of the great differences in the habitats, $d_f$=7/4 was general.

The same fractal dimension was observable not only in the snapshot hulls (sn), but also in the long-term averages (lt). According to this result, averaging did not create a more round-shaped hull, only decreased the width of the hull (see later), preserving its fractal structure. Furthermore, the fractal dimension did not change when we varied the occupancy threshold (*b*) at the delineation of the lt hull (**Fig. 4**), in spite of the fact that the hull shifted along the gradient, and its width also changed (**Fig. 2.b**).

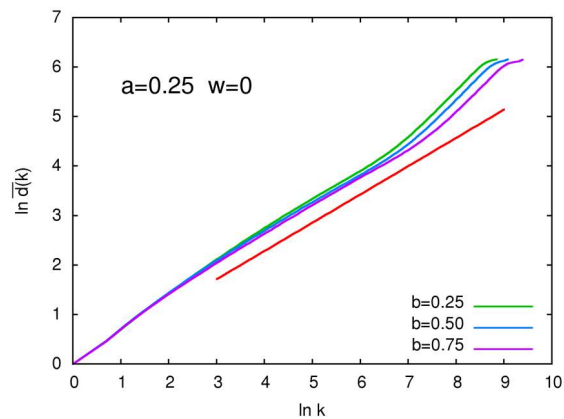

**Fig. 4.** *Estimation of the fractal dimension of the hull by the same method as in Fig. 3. The curves show long-term averages in the RS model at a system size L=1024. They differ in the arbitrary threshold value (b) which separates the occupied vs. unoccupied state. The straight line has a slope 4/7, which corresponds to a fractal dimension 7/4.*

The data plotted in **Figs. 3** and **4** deviate from the straight lines at low and high *k* because of finite size effects. At low *k*, the deviation is caused by the finite size of the lattice cells, due to which the spatial resolution cannot be infinitely fine, as should be in an ideal fractal. At high values of *k*, on the other hand, the deviation is caused by the environmental gradient, which confines the hull within a zone of width $v$ in the *x* direction. This width is determined by the condition that, at the edges of this zone, the characteristic linear size of islands or empty patches is comparable with the distance to the average position of the hull.
This yields a scaling law of the width in terms of the gradient (see Equation 5 below), in which the scaling exponent φ=4/7 can be expressed by the spatial correlation length's exponent of ordinary (homogeneous) percolation, $v_\perp$=4/3, via the general relationship $\varphi = v_\perp/(1 + v_\perp)$ (Sapoval et al. 1985). The width $v$ thus increases sublinearly with L (see Equation 5 below); therefore, in large systems it remains much smaller than the system size. The curve $\overline{d(k)}$ starts to deviate from the power law at the value of *k* at which $\overline{d(k)}$ becomes comparable with $v$. Well beyond this value of *k*, the hull appears as a one-dimensional object; thus, the slope of the curve gradually changes from 4/7 to 1. The range of *k* within which the hull can be considered a fractal is wider in larger systems. This is illustrated in **Fig. 3** at system sizes *L*=128, 256, 512, 1024, and 2048.

The scaling law concerning the fractal dimension (Eqn 3) is not the only one that can be expected. On the basis of theoretical considerations (Gastner et al. 2009), the length ($u$) and width ($v$) of the hull should also obey characteristic scaling laws,

$u(g) \propto g^{-\omega}$,     Eqn 4
$v(g) \propto g^{-\varphi}$,     Eqn 5

where $g$ is the environmental gradient. $\omega$ and $\varphi$ are universal exponents. Their values for the two-dimensional gradient percolation are $\omega=3/7$ and $\varphi=4/7$. The validity of these scaling laws has been confirmed by numerical simulations for the contact process in ES environments (Gastner et al. 2009). Our present study is the first to demonstrate the validity of these laws in RS environments as well (**Fig. 5**). It is remarkable that the exponents remain the same in different environments, varying the abundance ($a$) and penetrability ($w$) of the obstacles in the RS.

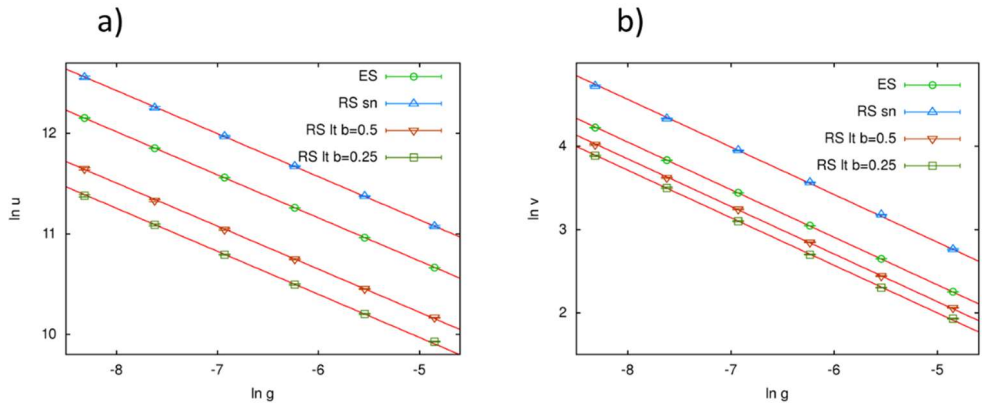

*Fig. 5.* Dependence of the **a)** length and **b)** width of the hull on the environmental gradient g in the RS model. The slopes of the straight lines are **a)** 4/7 and **b)** 3/7, according to the scaling laws expressed by Eqns 4 and 5, respectively. The size of the lattice in the y direction was $L_y=4096$ in every case, while the size in the x direction varied with g, according to $L_x=1/g$.

Discussion

In this work, we considered a metapopulation model on a two-dimensional lattice, in the presence of an environmental gradient and fine-scale heterogeneities. We partitioned the range margin into a connected and a fragmented part. According to our numerical results, the hull of the connected part (i.e., of the 'mainland') is a fractal, and has the same fractal dimension (7/4) as the traditional percolation hull under a broad variety of conditions. The fractal dimension, as well as the exponents characterizing the scaling of the width and the length of the hull were found to be universal irrespective of

– the presence vs. absence of fine-grained obstacles (RS vs. ES),
– the abundance of the obstacles ($a$),
– penetrability of the obstacles ($w$),
– the duration of the observation (snapshot vs. long-term),
– and, in case of long-term observations, the threshold above which we considered a site occupied ($b$).

The question arises how to explain this generality of the fractal dimension of the hull. The explanation is based on percolation theory. $d_f=7/4$ has been found in ordinary site percolation (Voss 1984; Saleur and Duplantier 1987; Stauffer and Aharony 1994), and in gradient percolation (Sapoval et al. 1985; Gastner et al. 2009). Nevertheless, $d_f$=7/4 in the ES and RS models is not self-evident.

In the ordinary and gradient percolation models, the states of the sites are uncorrelated. In contrast, in the ES and RS models they are correlated, since colonizations do not occur uniformly on arbitrary sites, but only on neighboring sites. Thus, the contact process produces a clumped pattern (**Figs. 1.c and d**). Let us define the correlation length ξ as the characteristic distance beyond which the occupancies can be considered statistically independent. As λ changes in the $x$ direction, ξ also changes (Oborny et al. 2009). It is well known that in the gradient free contact process ($g = 0$) there is a sharp extinction threshold ($\lambda_c$) at which the population becomes extinct in the infinite system. Furthermore, approaching this threshold, the correlation length ξ increases to infinity [see Marro et al. (2005); Henkel et al. (2008); Ódor (2008)]. In the gradient contact process ($g > 0$) let us refer to the coordinate $x_c$ where $\lambda(x_c) = \lambda_c$ as the *extinction limit*. As $x$ approaches the extinction limit, the local correlation length ξ thus increases. Why does this correlation not alter the percolation transition, i.e., why can we observe its characteristic exponents at the hull? The reason is that the hull is at higher $x$ values than the extinction limit, within the densely populated region (**Fig. 2.a**). Here, the correlation length is finite. At a length scale > ξ, the occupancies can be considered uncorrelated, and thus the scaling laws known from standard percolation models apply.

This conclusion can also be illuminated by the following argument, which is standard in the statistical physics of phase transitions. Let us assume that a coarse-graining of the system is performed by dividing the lattice into cells of linear size ξ, and, in the case of the map of long-term occupancies, an averaging within each cell is performed; while, in the case of a snapshot, a majority rule is applied to obtain discrete (0/1) occupancies of the cells. In this way, one obtains a coarse-grained system with practically uncorrelated occupancies of the cells. Thus, applying a further discretization on the map of long-term occupancies after coarse-graining, the resulting state does not essentially differ from that of the traditional gradient percolation model. As a consequence, the properties of the hull, when viewed on scales well beyond ξ, must be the same as those of the gradient percolation hull. Thus, the exponents appearing in the various power-laws are expected to be unaltered, and the details of the spatial pattern within the scale ξ affect, at most, the prefactors in the power laws.

The scaling laws found in this work are valid in the limit $g \to 0$, which is equivalent to $L \to \infty$, owing to the relation $g=1/L$ in our model. For small system sizes, corrections may be needed to the scaling laws, because the extinction limit may get so close to the percolation limit that the hull's shape gets distorted; therefore, a deviation from the asymptotic value of the exponent $d_f$=7/4 can be expected. **Fig. 3** demonstrates, however, that this kind of finite-size effect is negligible even in rather small systems, at $L$=128. The datapoints fitted well to the straight line, indicating $d_f$=7/4 in a considerable range of $k$.

Let us discuss the validity of the results for other possible implementations of the environmental gradient. Throughout the paper, we used a constant extinction rate and a linearly varying colonization rate, $c(x) = gx$, so that the control parameter, $\lambda(x)$ =c(x)/e, varies strictly linearly in the whole range of $x$. We could, however, have equally well chosen a constant colonization rate, $c$, and a constant gradient in $e$, with $e(x) = gx$. In that case, although the control parameter $\lambda(x)$ =c/e(x) is not a linear function of $x$, it can still be expanded in a Taylor series around the average position of the hull, $x_h$, as $\lambda(x) = \lambda(x_h) - \frac{[\lambda(x_h)]^2 g}{c}(x - x_h) + O([g(x - x_h)]^2)$. For a small gradient, the higher order terms are negligible near the linear one, and we arrive at the same linear dependence as in the previous case. The fractal properties of the hull are therefore expected to be universal also with respect to the particular implementation of the gradient (see also Gastner et al. 2012).

There are further noteworthy parameters, which influence the average position and width of the hull, and thus, influence whether it reaches to the extinction limit. First of all, the step length ($s$) at the delineation of the clusters is an arbitrary choice. The larger is $s$, the closer the hull gets to the extinction limit. Secondly, in the study of long-term averages, the choice of $b$ can also influence the position and width of the hull (**Fig. 2.b**). Choosing a larger $b$ increases the chance that the hull is not distorted.

It is interesting to consider that the fractal structure of the hull remains invariant in spite of shifting it along $x$ by varying $s$ (Gastner et al. 2009) or $b$ (in the present study). This indicates the existence of a considerable *fractal zone* within the range margin. The existence of this zone gives some freedom in choosing the parameter values at the detection of the hull. This is particularly useful when analyzing real-life data. A typical example is analyzing a satellite image which shows the occupancy of space by a species or vegetation type along a gradient [e.g., Milne et al. (1996); Gastner et al. (2009); Saravia et al. (2018)]. The first step is to mark out the largest percolation cluster using a given neighborhood size $s$. The next one is to delineate the hull with the same $s$. The value of $s$ is a matter of choice within reasonable limits, so that the hull should be within the margin, but sufficiently far from the extinction limit to avoid distortion. It is worth checking whether this fractal dimension is 7/4. Obtaining 7/4 confirms that the hull has been identified correctly, and there is no hidden environmental factor in the background which would invalidate our assumptions about the system, e.g., a sudden change in the environmental conditions instead of a gradient. It should be noted, however, that we have tested various density profiles at the hull, including cases in which the probability of occupation changed according to a step function, and these did not change the fractal dimension 7/4 (Gastner and Oborny 2012).

After checking the fractal dimension, we propose to mark the mean position of the hull along $x$ as the actual range edge. On those terrains in which the direction of the gradient is changing, we suggest estimating the mean in a sliding window. Theoretical considerations suggest that delineating the hull, i.e., the 'connectivity limit', is statistically more reliable that delineating the extinction limit (Oborny 2018). Therefore, we propose focusing on the hull when monitoring range shifts. The universal nature of the scaling relations at the hull, revealed by theoretical models, suggests that range shifts in different geographic regions, even in different species can be compared directly, as the fine-scale details of the pattern-generating processes become irrelevant at the hull, and produce the same fractal structure. Our present results broaden the scope of the applicability of the method to RS environments.

## Outlook

Species range edges can be influenced by other species, for example, by competitors. This underlines the utility of delineating the range edge at the hull, where the species occurs in relatively high density: the population's dynamics is probably less disturbed by external influences at the hull than at the extinction edge. In general, the study of multi-species edges with positive and negative feedback loops (including those which act through modifying the environment) is an important matter for future research.

A considerable value of percolation-based approaches is that they deepen our understanding of connectivity within the range of a species, which is a key to many ecological and evolutionary processes. For example, connectivity can affect gene flow; and thus, can influence the pace and direction of microevolution (Geber 2008; Kunin et al. 2009; Kubisch et al. 2014). Low connectivity, and thus, limited gene flow along an environmental gradient may be beneficial or detrimental for the survival of a (meta)population, depending on the species and key environmental factors. On the positive side, low connectivity can facilitate local adaptation (Kubisch et al. 2014). Thus, it can increase genetic diversity on a broad, geographic scale (Gaston 2009; Kunin et al. 2009). On the negative side, it may lead to the loss of diversity on the local scale, especially within small islands of occurrence (Gaston 2009; Kunin et al. 2009). Low gene flow may also prevent the spreading of those mutations which arise further away from their optimum place along the gradient (Turner and Wong 2010). These factors become particularly important when the climate is changing. In this case, survival of a species at its range margin may hinge on its ability to colonize new locations, and to adapt to the environment locally (Brown et al. 1996; Geber 2008; Kubisch et al. 2014). Thus, the study of range margins in terms of connectivity can provide important pieces of information about the responses of species to climate change.

Connectivity of the occupied sites determines not only the flow of genes, but more generally, the flow of material and energy. For example, the connected vs. fragmented occurrence of shrubs in a shrubland or of trees in a woodland can significantly influence the evaporation and diffusion of water (Vincenot et al. 2016; Meron 2016; Ilstedt et al. 2016), and the spreading of pathogens (Orozco-Fuentes et al. 2019) and of fire (Abades et al. 2014). The density of trees has been shown to influence the access to food of the chipmunk *Eutamias umbrinus* against it more aggressive competior, *Eutamias dorsalis*, as *E. umbrinus* can use the trees as shelters (Brown 1971). Connectivity in a population of an insect-pollinated plant can significantly influence the pollinator's foraging behavior (Bernhardt et al. 2008). When connectivity changes along an environmental gradient, these behaviors can also be hypothesized to change.

It is particularly interesting to consider the pattern of occupied sites from the aspect of another species which uses the present species as a habitat [see, for example, a review by Saravia et al. (2018) about global forest fragmentation]; or, conversely, requires unoccupied sites for spreading (Solé et al. 2005). Habitat connectivity is one of the central topics in landscape ecology (Gardner et al. 1987; With et al. 1997; Fahrig 2003; Oborny et al. 2007; Solé 2011). The focus is primarily on those cases in which the patch pattern is caused by abiotic factors (e.g. topography, hydrology) or by human activities (land use). The present study highlights the importance of studying connectivity in those cases in which the patch pattern is generated (at least partly) biotically, through metapopulation dynamics. In our ES, metapopulation dynamics was the only pattern-generating factor (modelled by a contact process). The RS was a mixed case, as metapopulation dynamics met a pre-existing pattern of good and bad sites. In both cases, the mainland-island structure emerged spontaneously. This pattern was dynamically changing, i.e., islands could merge with the mainland, and new islands were born by splitting. The study of these dynamically changing patterns is an exciting challenge for landscape ecology [c.f. Kun et al. (2019)].

In conclusion, the hull is a demarcation line between the core and periphery of a metapopulation in terms of connectivity. It has a robust fractal structure, which is insensitive to the details of delineation. We suggest marking the edge of a species' range at the mean position of the hull. This allows for detecting range shifts. The universal features of the hull make it possible to compare different species and/or geographic locations, even if the limiting factors are different. Several papers have called for a unified theoretical view of range margins [e.g., Brown et al. (1996); Maurer and Taper (2002); Holt and Keitt (2005); Antonovics et al. (2006); Gaston (2009); Kubisch et al. (2014)]. We believe that utilizing the universalities that emerge at critical phase transitions, independently of the details of the system, is an important step in this direction.

Acknowledgements
We are grateful to Michael Gastner, György Szabó and Géza Ódor for helpful comments on the manuscript.

Funding

This work was supported by the Hungarian Scientific Research Fund under grant no. K128989 and K124438, and a GINOP grant no. 2.3.2–15–2016–00019.

References

Abades SR, Gaxiola A, Marquet PA (2014) Fire, percolation thresholds and the savanna forest transition: a neutral model approach. J Ecol 102:1386–1393. https://doi.org/10.1111/1365-2745.12321

Antonovics J, McKane AJ, Newman TJ (2006) Spatiotemporal dynamics in marginal populations. Am Nat 167:16–27. https://doi.org/10.1086/498539

Batty M, Longley P (1994) Fractal cities: A geometry of form and function. Academic Press Professional, Inc., San Diego, CA, USA

Bernhardt CE, Mitchell RJ, Michaels HJ (2008) Effects of population size and density on pollinator visitation, pollinator behavior, and pollen tube abundance in Lupinus perennis. Int J Plant Sci 169:944–953. https://doi.org/10.1086/589698

Best AS, Johst K, Münkemüller T, Travis JMJ (2007) Which species will succesfully track climate change? The influence of intraspecific competition and density dependent dispersal on range shifting dynamics. Oikos 116:1531–1539. https://doi.org/10.1111/j.0030-1299.2007.16047.x

Brown JH (1971) Mechanisms of competitive exclusion between two species of chipmunks. Ecology 52:305–311. https://doi.org/10.2307/1934589

Brown JH, Stevens GC, Kaufman DM (1996) The geographic range: size, shape, boundaries, and internal structure. Annu Rev Ecol Syst 27:597–623. https://doi.org/10.1146/annurev.ecolsys.27.1.597

Bunde ASH (1991) Fractals and disordered systems. Springer-Verlag, Berlin

Eppinga MB, Pucko CA, Baudena M, et al (2013) A new method to infer vegetation boundary movement from 'snapshot' data. Ecography 36:622–635. https://doi.org/10.1111/j.1600-0587.2012.07753.x

Fahrig L (2003) Effects of habitat fragmentation on biodiversity. Annu Rev Ecol Evol Syst 34:487–515. https://doi.org/10.1146/annurev.ecolsys.34.011802.132419

Gardner RH, Milne BT, Turnei MG, O'Neill RV (1987) Neutral models for the analysis of broad-scale landscape pattern. Landsc Ecol 1:19–28. https://doi.org/10.1007/BF02275262

Gastner MT, Oborny B (2012) The geometry of percolation fronts in two-dimensional lattices with spatially varying densities. New J Phys 14:103019. https://doi.org/10.1088/1367-2630/14/10/103019

Gastner MT, Oborny B, Zimmermann DK, et al (2009) Transition from connected to fragmented vegetation across an environmental gradient: Scaling laws in ecotone geometry. Am Nat 174:E23–E39. https://doi.org/10.1086/599292

Gaston KJ (2009) Geographic range limits: achieving synthesis. Proc R Soc B Biol Sci 276:1395–1406. https://doi.org/10.1098/rspb.2008.1480

Geber MA (2008) To the edge: studies of species' range limits. New Phytol 178:228–230. https://doi.org/10.1111/j.1469-8137.2008.02414.x

Harris TE (1974) Contact interactions on a lattice. Ann Probab 2:969–988

Henkel M, Hinrichsen H, Lübeck S (2008) Non-equilibrium phase transitions: Volume 1: Absorbing phase transitions. Springer Netherlands

Holt RD, Keitt TH (2000) Alternative causes for range limits: a metapopulation perspective. Ecol Lett 3:41–47. https://doi.org/10.1046/j.1461-0248.2000.00116.x

Holt RD, Keitt TH (2005) Species' borders: a unifying theme in ecology. Oikos 108:3–6. https://doi.org/10.1111/j.0030-1299.2005.13145.x

Holt RD, Keitt TH, Lewis MA, et al (2005) Theoretical models of species' borders: single species approaches. Oikos 108:18–27. https://doi.org/10.1111/j.0030-1299.2005.13147.x

Ilstedt U, Bargués Tobella A, Bazié HR, et al (2016) Intermediate tree cover can maximize groundwater recharge in the seasonally dry tropics. Sci Rep 6:21930. https://doi.org/10.1038/srep21930

Kubisch A, Holt RD, Poethke H-J, Fronhofer EA (2014) Where am I and why? Synthesizing range biology and the eco-evolutionary dynamics of dispersal. Oikos 123:5–22. https://doi.org/10.1111/j.1600-0706.2013.00706.x

Kun Á, Oborny B, Dieckmann U (2019) Five main phases of landscape degradation revealed by a dynamic mesoscale model analysing the splitting, shrinking, and disappearing of habitat patches. Sci Rep 9:11149. https://doi.org/10.1038/s41598-019-47497-7

Kunin WE, Vergeer P, Kenta T, et al (2009) Variation at range margins across multiple spatial scales: environmental temperature, population genetics and metabolomic phenotype. Proc Biol Sci 276:1495–1506. https://doi.org/10.1098/rspb.2008.1767

Lennon JJ, Turner JRG, Connell D (1997) A metapopulation model of species boundaries. Oikos 78:486–502. https://doi.org/10.2307/3545610

Li B-L (2001) Applications of Fractal Geometry and Percolation Theory to Landscape Analysis and Assessments. In: Jensen ME, Bourgeron PS (eds) A Guidebook for Integrated Ecological Assessments. Springer, New York, NY, pp 200–210

Loehle C, Li B-L, Sundell RC (1996) Forest spread and phase transitions at forest-prairie ecotones in Kansas, U.S.A. Landsc Ecol 11:225–235. https://doi.org/10.1007/BF02071813

Marro J, Dickman R (2005) Nonequilibrium phase transitions in lattice models. Cambridge University Press, Cambridge

Maurer BA, Taper ML (2002) Connecting geographical distributions with population processes. Ecol Lett 5:223–231. https://doi.org/10.1046/j.1461-0248.2002.00308.x

Meron E (2016) Pattern formation – A missing link in the study of ecosystem response to environmental changes. Math Biosci 271:1–18. https://doi.org/10.1016/j.mbs.2015.10.015

Milne BT, Johnson AR, Keitt TH, et al (1996) Detection of critical densities associated with piñon-juniper woodland ecotones. Ecology 77:805–821. https://doi.org/10.2307/2265503

Moreira AG, Dickman R (1996) Critical dynamics of the contact process with quenched disorder. Physical Review E 54:R3090-R3093. https://doi.org/10.1103/PhysRevE.54.R3090

Mustin K, Benton TG, Dytham C, Travis JMJ (2009) The dynamics of climate-induced range shifting: perspectives from simulation modelling. Oikos 118:131–137. https://doi.org/10.1111/j.1600-0706.2008.17025.x

Oborny B (2018) Scaling laws in the fine-scale structure of range margins. Mathematics 6:315. https://doi.org/10.3390/math6120315

Oborny B, Szabó G, Meszéna G (2007) Survival of species in patchy landscapes : percolation in space and time. In: Storch D, Marquet P, Brown J (eds) Scaling biodiversity. Cambridge University Press, Cambridge, pp 409–440

Oborny B, Vukov J, Csányi G, Meszéna G (2009) Metapopulation dynamics across gradients – the relation between colonization and extinction in shaping the range edge. Oikos 118:1453–1460. https://doi.org/10.1111/j.1600-0706.2009.17158.x

Ódor G (2008) Universality In nonequilibrium lattice systems: Theoretical foundations. World Scientific Pub Co Inc, Singapore ; Hackensack, NJ

Orozco-Fuentes S, Griffiths G, Holmes MJ, et al (2019) Early warning signals in plant disease outbreaks. Ecol Model 393:12–19. https://doi.org/10.1016/j.ecolmodel.2018.11.003

Saleur H, Duplantier B (1987) Exact determination of the percolation hull exponent in two dimensions. Phys Rev Lett 58:2325–2328. https://doi.org/10.1103/PhysRevLett.58.2325

Sapoval B, Rosso M, Gouyet JF (1985) The fractal nature of a diffusion front and the relation to percolation. J Phys Lett 46:149–156. https://doi.org/10.1051/jphyslet:01985004604014900

Saravia LA, Doyle SR, Bond-Lamberty B (2018) Power laws and critical fragmentation in global forests. Sci Rep 8:1–12. https://doi.org/10.1038/s41598-018-36120-w

Solé RV (2011) Phase Transitions, Princeton University Press, Princeton, NJ

Solé RV, Bartumeus F, Gamarra JG (2005) Gap percolation in rainforests. Oikos, 110:177-185, 10.1111/j.0030-1299.2005.13843.x

Stauffer D, Aharony A (1994) Introduction to percolation theory: Second Edition. Taylor and Francis

Tejo M, Niklitschek-Soto S, Vásquez C, Marquet PA (2017) Single species dynamics under climate change. Theor Ecol 10:181–193. https://doi.org/10.1007/s12080-016-0321-0

Travis JMJ (2003) Climate change and habitat destruction: a deadly anthropogenic cocktail. Proc R Soc B Biol Sci 270:467–473. https://doi.org/10.1098/rspb.2002.2246

Turner JRG, Wong HY (2010) Why do species have a skin? Investigating mutational constraint with a fundamental population model. Biol J Linn Soc 101:213–227. https://doi.org/10.1111/j.1095-8312.2010.01475.x

Vincenot CE, Carteni F, Mazzoleni S, et al (2016) Spatial self-organization of vegetation subject to climatic stress: Insights from a system dynamics - individual-based hybrid model. Front Plant Sci 7:. https://doi.org/10.3389/fpls.2016.00636

Voss RF (1984) The fractal dimension of percolation cluster hulls. J Phys Math Gen 17:L373–L377. https://doi.org/10.1088/0305-4470/17/7/001

Vucetich JA, Waite TA (2003) Spatial patterns of demography and genetic processes across the species' range: Null hypotheses for landscape conservation genetics. Conserv Genet 4:639–645. https://doi.org/10.1023/A:1025671831349

Wilson WG, Nisbet RM, Ross AH, et al (1996) Abrupt population changes along smooth environmental gradients. Bull Math Biol 58:907. https://doi.org/10.1007/BF02459489

With KA, Gardner RH, Turner MG (1997) Landscape connectivity and population distributions in heterogeneous environments. Oikos 78:151–169. https://doi.org/10.2307/3545811

Zeng Y, Malanson GP (2006) Endogenous fractal dynamics at Alpine treeline ecotones. Geogr Anal 38:271–287. https://doi.org/10.1111/j.1538-4632.2006.00686.x